\documentclass[]{pasj02} 
\usepackage[switch,mathlines]{lineno} % add line number to manuscript
\usepackage{lscape}
\usepackage{graphics}
\usepackage{comment}
\usepackage{adjustbox}
\usepackage{url}
\usepackage{bm}
\usepackage{amsmath}
\usepackage{mathtools}

\newcommand{\KH}[1]{\textcolor{black}{#1}}
\newcommand{\MM}[1]{{#1}}%{\textcolor{red}{#1}}
\usepackage{ulem}

\jyear{2026}
\Received{}%{yyyy/mm/dd}
\Accepted{}%{yyyy/mm/dd}
\begin{document} 

%\title{ Cusp-to-Core Transition Phase on Galaxy Mass Scales }
\title{ Cusp-to-Core Transition of Dark Matter Halos across Galaxy Mass Scales }

%%% begin:list of authors
% Do NOT capitalize all letters in "textsc".
\author{
 Kohei \textsc{Hayashi}\altaffilmark{1,2,3}\altemailmark\orcid{0000-0002-8758-8139} \email{khayashi@sendai-nct.ac.jp},
 Yuka \textsc{Kaneda}\altaffilmark{4}, \orcid{0009-0002-3547-9472} 
 Masao \textsc{Mori}\altaffilmark{5}, \orcid{0000-0002-2883-8943}
 and
 Michi \textsc{Shinozaki}\altaffilmark{4}
}
\altaffiltext{1}{National Institute of Technology, Sendai College, 4-16-1 Ayashi-Chuo, Sendai, Japan}
\altaffiltext{2}{Astronomical Institute, Tohoku University, 6-3 Aoba, Sendai, Japan}
\altaffiltext{3}{ICRR, The University of Tokyo, 5-1-5 Kashiwanoha, Kashiwa, Japan}
\altaffiltext{4}{Graduate School of Science and Technology, University of Tsukuba, 1-1-1 Tennodai, Tsukuba, Ibaraki, 305-8577, Japan}
\altaffiltext{5}{Center for Computational Sciences , University of Tsukuba, 1-1-1 Tennodai, Tsukuba, Ibaraki, 305-8577, Japan}

%\footnotetext[$\dag$]{Present address: ....}

%%% end:list of authors

%% !!! Select 3 to 5 words from PASJ's key words !!! 
%% List of Key Words: https://academic.oup.com/pasj/pages/Pasj_Keywords 
%% "\KeyWords{ }" always has to be placed before ``\maketitle'' 

\maketitle

\begin{abstract}
We investigate the diversity of dark matter (DM) density profiles in a large sample of late-type galaxies from the SPARC database, with the goal of testing whether a cusp-to-core transition occurs across galaxy mass scales.
We perform Bayesian fits \MM{to} high-quality rotation curves using flexible halo models that allow for variations in \MM{the} inner slopes of DM density profiles. 
We quantify the central dark matter structure using the surface density within the inner region of the halo, defined as $\Sigma_{\rm DM}(<0.01r_{V_{\rm max}})$, and compare the SPARC galaxies with Milky Way dwarf satellites as well as galaxy groups and clusters.
\MM{Our results reveal significant diversity in the inner density slopes of SPARC galaxies, ranging from steep cusps to shallow cores, and show that many of them lie below the cuspy profiles predicted by the cold dark matter model, consistent with core-like structures.} 
In contrast, both lower-mass dwarf galaxies and higher-mass galaxy clusters tend to follow the \MM{cuspy DM halos. These findings suggest} that baryonic feedback may induce a cusp-to-core transition in Milky Way–mass galaxies, as predicted by hydrodynamical simulations. 
However, observational limitations and modeling uncertainties still prevent a definitive conclusion. 
This study provides new empirical insights into the halo mass-dependent nature of DM inner structures and the role of baryonic processes in shaping them.
\end{abstract}

%\pagewiselinenumbers 

%%%%%%%%%%%%%%%%%%%%%%%%%%%%%%
%%%%%%%%% Section 1 %%%%%%%%%%
%%%%%%%%%%%%%%%%%%%%%%%%%%%%%%

\section{Introduction}
The $\Lambda$ cold dark matter ($\Lambda$CDM) model has been remarkably successful in explaining the large-scale structure of the Universe, the cosmic microwave background radiation~(e.g., \cite{2011ApJS..192...18K,2020AA...641A...6P}), and the distribution of galaxies~(e.g., \cite{2006PhRvD..74l3507T,2025JCAP...02..021A}). 
Despite its successes on these scales, significant challenges remain when the model is applied to galactic and sub-galactic scales~(see, \cite{2017ARAA..55..343B} for a review). 

One of the most notable is the so-called ``cusp–core'' problem. 
Pure dark matter simulations based on the $\Lambda$CDM models predict that dark matter halos should form a universal dark matter density profile with a steep cusp at the center (e.g., \cite{1991ApJ...378..496D,1994Natur.370..629M,1997ApJ...490..493N,1997ApJ...477L...9F,2013ApJ...767..146I}).
In contrast, observations of dwarf spheroidal galaxies (dSphs)  and low surface brightness galaxies (LSBs) often suggest shallower or constant central dark matter density (e.g., \cite{1995ApJ...447L..25B,2001ApJ...552L..23D,2007ApJ...663..948G,2008AJ....136.2761O,2010AdAst2010E...5D}).
More recently, dynamical studies of Galactic dSphs and the rotation curves of late-type galaxies (including LSBs) have revealed diversity in inner dark matter densities, ranging from cusps to cores~(e.g., \cite{2015MNRAS.452.3650O,2019MNRAS.490..231K,2019MNRAS.484.1401R,2020ApJ...904...45H,2023ApJ...953..185H,2016AJ....152..157L})\KH{, although part of this diversity may be affected by systematic uncertainties in dwarf galaxy rotation curve measurements (e.g., \cite{2023MNRAS.522.3318D}).}
This \KH{may} represents a new issue for the $\Lambda$CDM model, commonly referred to as the “diversity problem” rather than simply the “cusp–core” problem.

Two main approaches have been proposed to address this discrepancy. 
The first involves baryonic feedback processes, such as supernova-driven gas outflows~(e.g., \cite{1996MNRAS.283L..72N,2002MNRAS.333..299G,2005MNRAS.356..107R,2011ApJ...736L...2O,2012MNRAS.424.1275B,2012MNRAS.421.3464P,2014ApJ...793...46O}) and interactions between gas clumps and dark matter~(e.g., \cite{2001ApJ...560..636E,2011MNRAS.418.2527I,2015MNRAS.446.1820N}). 
In the case of supernova feedback, recent high-resolution simulations have shown that the efficiency of core formation depends sensitively on the stellar-to-halo mass ratio and the star formation history (e.g., \cite{2014MNRAS.441.2986D,2014MNRAS.437..415D,2015MNRAS.454.2092O,2016MNRAS.456.3542T,2017MNRAS.471.3547F,2020MNRAS.497.2393L}).
While several studies, particularly those based on the rotation curves of late-type galaxies, have reported that state-of-the-art $\Lambda$CDM-based hydrodynamical simulations still fall short of fully reproducing the observed diversity in inner dark matter densities, even when baryonic feedback is taken into account (e.g., \cite{2015MNRAS.452.3650O,2020MNRAS.495...58S}), other studies have recently argued that some observed rotation curves may significantly deviate from true circular velocity curves due to a combination of non-circular motions, geometrically thick gas disks, and dynamical disequilibrium~(\cite{2023MNRAS.521.1316R,2024arXiv240416247S}).
%The effectiveness of baryonic feedback in forming cores depends strongly on factors such as stellar mass, star formation history, and the gas density threshold adopted in simulations.

The second approach involves alternative dark matter models, such as self-interacting dark matter (SIDM;  \cite{1992ApJ...398...43C,2000PhRvL..84.3760S,2016PhRvL.116d1302K}) and fuzzy dark matter (FDM; \cite{1978PhRvL..40..223W,2000PhRvL..85.1158H,2016PhR...643....1M,2021AnARv..29....7F}).
These models predict the formation of central cores through mechanisms rooted in their fundamental particle properties, independent of baryonic feedback.
For instance, SIDM leads to core formation via elastic scattering between dark matter particles~(\cite{2013MNRAS.430...81R,2013MNRAS.431L..20Z,2020ApJ...896..112N}), while FDM induces core-like structures due to quantum pressure arising from the ultra-light bosonic nature of the particles~(\cite{2014NatPh..10..496S,2016PhRvD..94d3513S,2017MNRAS.471.4559M,2022MNRAS.511..943C}).
In particular, SIDM models coupled with baryonic gravitational potential (e.g., \cite{2017PhRvL.119k1102K,2019PhRvX...9c1020R}) or incorporating gravothermal core-collapse mechanisms (e.g., \cite{2020PhRvD.101f3009N,2021MNRAS.503..920C,2025PhRvD.111j3041R}) have been shown to reproduce the observed diversity in dark matter densities of dwarf satellites and late-type galaxies.
As a result, alternative dark matter models are becoming increasingly attractive as potential explanations for discrepancies between $\Lambda$CDM predictions and observational data.

From an observational perspective, understanding the scaling relations among dark matter halos plays a crucial role in investigating the nature of dark matter.
These relations describe empirical correlations between various halo properties, such as mass, characteristic size, concentration, and density. Over the past decades, these scaling laws have been extensively studied, as they provide important constraints on models of galaxy formation and evolution.
Dynamical analyses of galaxies, including dSph and late-type galaxies, have revealed that the central surface density of dark matter halos, defined as the product of the central density and the core radius assuming cored dark matter density profiles, remains nearly constant across a luminosity range spanning approximately 14 magnitudes~(e.g., \cite{2004IAUS..220..377K,2008MNRAS.383..297S,2009MNRAS.397.1169D,2012MNRAS.420.2034S,2016ApJ...817...84K}).
In contrast, several studies have reported a mild dependence of the central surface density on galaxy scale, consistent with predictions from $\Lambda$CDM $N$-body simulations \MM{(e.g., \cite{2010PhRvL.104s1301B,2014MNRAS.440L..71O,2017ApJ...843...97H,2019MNRAS.482.5106L}).}
%----------------------
More recently, \citet{2024PASJ...76.1026K} proposed a new scaling relation for the central surface density of dark matter halos that incorporates the cusp-to-core transition effect. 
While this relation provides a promising framework to test the existence of such transitions, their comparison was limited to dark matter halos whose density profiles were assumed to be fixed, rather than allowing for variation in the inner slope.

As mentioned above, the dark matter halos of dSphs and late-type galaxies have been extensively studied because these systems are believed to be largely dominated by dark matter.
In previous studies for dSphs, their stellar kinematic data, primarily line-of-sight velocities, have typically been used to estimate their dark matter density profiles. 
Recently, thanks to precise stellar imaging and a long observational baseline, the Hubble Space Telescope (HST) has provided proper motions of individual stars in luminous dSphs~(\cite{2018NatAs...2..156M,2020AaA...633A..36M,2024ApJ...970....1V}).
By combining these with existing line-of-sight velocity data, it has become possible to obtain resolved three-dimensional velocity measurements for these galaxies.
Such data are analyzed using dynamical methods, including Jeans modeling, distribution function approaches, and orbit-based techniques~(see \cite{2013NewAR..57...52B, 2022NatAs...6..659B} for reviews).
Most of these studies parameterize the dark matter density profile using quantities including the scale density, scale radius, inner and outer slopes, and halo shape~(as detailed in Equation~\ref{eq:DMH}). These models are then fitted to the observed velocity dispersion profiles of the dSphs.

For late-type galaxies, the Spitzer Photometry and Accurate Rotation Curves (SPARC) survey has provided high-quality rotation curve data for 175 galaxies, enabling a comprehensive investigation of dark matter halo models as well as modified Newtonian dynamics~(MOND, \cite{1983ApJ...270..371M}) through homogeneous fitting procedures~(e.g., \cite{2016AJ....152..157L,2016PhRvL.117t1101M,2018AnA...615A...3L,2020ApJS..247...31L}).
Most previous studies of late-type galaxies, including those utilizing the SPARC database, have adopted fixed functional forms for dark matter density profiles (such as the Navarro–Frenk–White, pseudo-isothermal, Burkert, Einasto and so on) and evaluated their relative performance by comparing the cumulative distributions of reduced $\chi^2$ values.
While these models provide useful benchmarks, they may oversimplify the actual diversity in dark matter halo structure, potentially overlooking transitional cases between cusps and cores.
In addition, observational constraints, particularly for low-mass systems, remain limited by data quality, which introduces significant uncertainties into the inferred dark matter density distributions.
Importantly, as mentioned earlier, $\Lambda$CDM-based hydrodynamical simulations predict that baryonic feedback can induce a cusp-to-core transition at the mass scale of Milky Way–like galaxies. 
However, such a transition has yet to be clearly confirmed by observations.

In this study, we aim to advance our understanding of the cusp-to-core transition across galaxy mass scales by analyzing a large sample of galaxies from the SPARC database. 
We employ flexible dark matter halo models that allow for variations in both the inner slope and overall shape of the halo.
Using Bayesian inference techniques to fit the high-quality rotation curves, and by examining the central surface density of the dark matter halo, we test whether the data provide evidence for a cusp-to-core transition across different galaxy mass scales.

The structure of this paper is as follows.
Section~2 describes the data,  the modeling methodology and fitting procedures.
Section~3 presents the main results.
Section~4 shows the cusp-to-core transition phase using the 
\MM{dark matter (DM)} surface density.
Section~5 discusses possible evidence for the cusp-to-core transition phase and discrepancies between the observations and theoretical predictions.
Finally, the conclusions are summarized in Section~6.

%%%%%%%%%%%%%%%%%%%%%%%%%%%%%%
%%%%%%%%% Section 2 %%%%%%%%%%
%%%%%%%%%%%%%%%%%%%%%%%%%%%%%%
\section{Data, models, and method}\label{sec:2}
\subsection{The SPARC data}

The SPARC dataset\footnote{\url{http://astroweb.cwru.edu/SPARC/}} is a comprehensive catalog of 175 late-type galaxies, providing high-quality HI/H$\alpha$ rotation curves and near-infrared Spitzer photometry at $3.6\mu$m. This dataset is particularly useful for studying the distribution of 
%MM-------------------
\MM{DM} in galaxies because of its ability to trace the rotation velocity ($V_\mathrm{obs}$) out to large radii, offering strong constraints on the underlying dark matter density profiles.

The near-infrared photometry is a key feature of the SPARC dataset, as it significantly reduces the scatter in the stellar mass-to-light ratio, especially at $3.6\mu$m. 
This allows for more accurate modeling of the stellar disk's contribution to the galaxy's total mass and helps address the disk-halo degeneracy. The mass models for the stellar disk and, when applicable, the bulge, are derived by solving the Poisson equation for the observed surface brightness profiles, while the gas contribution is based on the HI surface density profile, scaled to include Helium.
The SPARC sample covers a broad range of galaxy properties, including \KH{luminosities ($\sim10^7$ to $\sim10^{12}$\,$L_\odot$) and surface brightness ($\sim 5$ to $\sim5000$\, $L_\odot\,\mathrm{pc}^{-2}$)}. This diversity makes the dataset ideal for probing various dark matter models and examining how the properties of dark matter halos correlate with the visible matter in galaxies.

Galaxy distances in the SPARC database are determined through a variety of methods, including Hubble flow, red giant branch tip magnitude, Cepheid period-luminosity relation, and Type Ia supernovae measurements. Disk inclinations are derived kinematically, and uncertainties in both distance and inclination are treated as nuisance parameters during model fitting.

These galaxies are particularly valuable for dark matter studies because their low baryonic content means that their rotation curves are dominated by dark matter across a significant portion of their radial extent. By using the SPARC dataset, we aim to extract detailed information about the dark matter density distribution and compare it with theoretical models.

Among these data, we impose selection requirements with quality index less than 3, which means we exclude the galaxy sample with low rotation-curve quality taken from \citet{2016AJ....152..157L}. 
Since we have 9 (10) free fitting parameters for disk (disk plus bulge) SPARC galaxies in this work, \MM{we also exclude galaxies with fewer than 9 data points in their observed rotation curves.}
Furthermore, we  select only the galaxies with disk inclination angles greater than $30^\circ$ to avoid using nearly face-on galaxies.
After this selection, we get 115 out of 175 galaxies.

\subsection{Dark matter density profiles}

In this paper, we adopt a generalized Hernquist profile \citep{1990ApJ...356..359H,1996MNRAS.278..488Z}, while considering axisymmetric dark matter halos in cylindrical coordinates~$(R,z)$:
\begin{eqnarray}
&& \rho_{\rm DM}(R,z) = \rho_0 \Bigl(\frac{m}{b_\mathrm{halo}} \Bigr)^{-\gamma}\Bigl[1+\Bigl(\frac{m}{b_\mathrm{halo}} \Bigr)^{\alpha}\Bigr]^{-\frac{\beta-\gamma}{\alpha}},
 \label{eq:DMH} \\
&& m^2=R^2+z^2/Q^2,
\label{eq:DMH2}
\end{eqnarray}
where $\rho_s$ and $r_s$ are the scale density and radius, respectively; $\alpha$ is
the sharpness parameter of the transition from the inner slope $\gamma$ to the outer slope $\beta$; and $Q$ is a constant axial ratio of a dark matter halo.
This model can cover a broad range of physically plausible dark matter profiles from the cusped Navarro–Frenk–White (NFW) profile to the cored Burkert profile.

Using the dark matter density profile, we calculate the circular velocity given by \KH{$V^2_\mathrm{circ,DM}(R)=R|-\nabla\Phi_\mathrm{DM}(R,0)|$}, where $\Phi_\mathrm{DM}$ is a gravitational potential from DM.
In a spheroidal system, we can obtain the gravitational force in a straightforward manner by following~\citet{2008gady.book.....B}. 
Using a new variable of integration, $\tau\equiv a_0^2(1-Q^2) [\sinh^2u_m - (1/(1-Q^2)-1)]$\KH{, where $a_0$ is any constant and $u_m$ is a spheroidal coordinate with the homoeoid labeled by $m$.\footnote{\KH{A spheroidal coordinates $(u, v)$ are defined by $R=\Delta\cosh u\sin v$, $z=\Delta\sinh u\cos v$ $(u\geq0,0\leq v\leq \pi)$, where $\Delta$ is a constant.}}}~(see Equation~(2.124) in \cite{2008gady.book.....B}), Equation~(\ref{eq:DMH2}) is transformed to
\begin{equation}
\frac{m^2}{a_0^2} = \frac{R^2}{\tau+a^2_0} + \frac{z^2}{\tau+Q^2a^2_0}.
\end{equation}
The gravitational \KH{field} is thus given in the form of one-dimensional integration:
\begin{equation}
\bm{g} = -\nabla\Phi_\mathrm{DM} = -\pi GQa_0\int^{\infty}_{0} d\tau \frac{\rho_\mathrm{DM}(m^2)\nabla m^2}{(\tau+a^2_0)\sqrt{\tau+Q^2a^2_0}},
\label{eq:grav_force}
\end{equation}
where
\begin{equation}
\nabla m^2 = 2a^2_0 \left( \frac{R}{\tau+a^2_0}\hat{\bf{e}}_R  + \frac{z}{\tau + Q^2a^2_0} \hat{\bf{e}}_z\right),
\end{equation}
and $(\hat{\bf{e}}_R,\hat{\bf{e}}_z)$ are unit vectors in the directions of $R$ and $z$.

To calculate the circular velocity, we utilize the radial component of the gravitational field from Equation~(\ref{eq:grav_force}),
\begin{equation}
g_R(R,z) = -\frac{\partial \Phi_\mathrm{DM}}{\partial R} = -2\pi G Qa^3_0R \int^\infty_0 d\tau \frac{\rho_\mathrm{DM}(m^2)}{(\tau+a^2_0)^2\sqrt{\tau+Q^2a^2_0}}.
\end{equation}
In this work, we estimate the circular velocity along the major axis, $z=0$, for simplicity and thus the $V_\mathrm{circ,DM}$ is written as,
\begin{eqnarray}
   V^2_\mathrm{circ,DM}(R) &=& -Rg_R(R,0) \nonumber \\
    &=& 4\pi GQ \int^R_0 dm \frac{m^2\rho_\mathrm{DM}(m^2)}{\sqrt{R^2-m^2(1-Q^2)}}.
\label{eq:v_cric} 
\end{eqnarray}
\KH{In summary, we have six free parameters of a dark matter halo $(\rho_0, b_\mathrm{halo}, Q, \alpha, \beta,\gamma)$.}

\begin{figure*}
    \includegraphics[width=.49\textwidth]{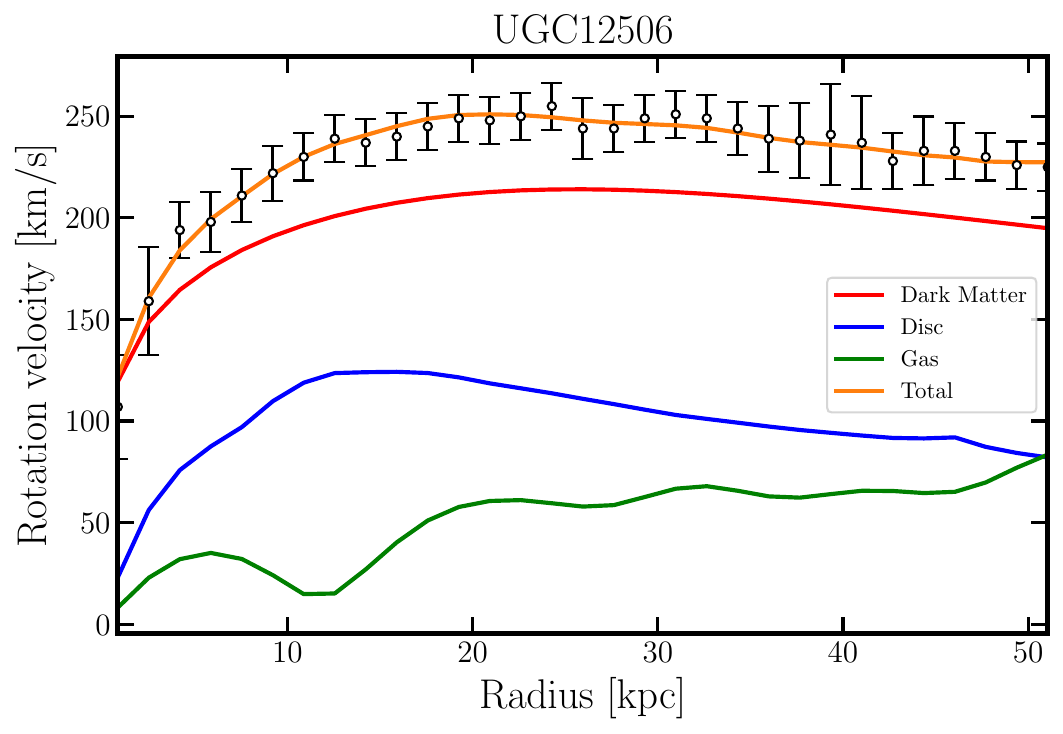}\hfill
    \includegraphics[width=.49\textwidth]{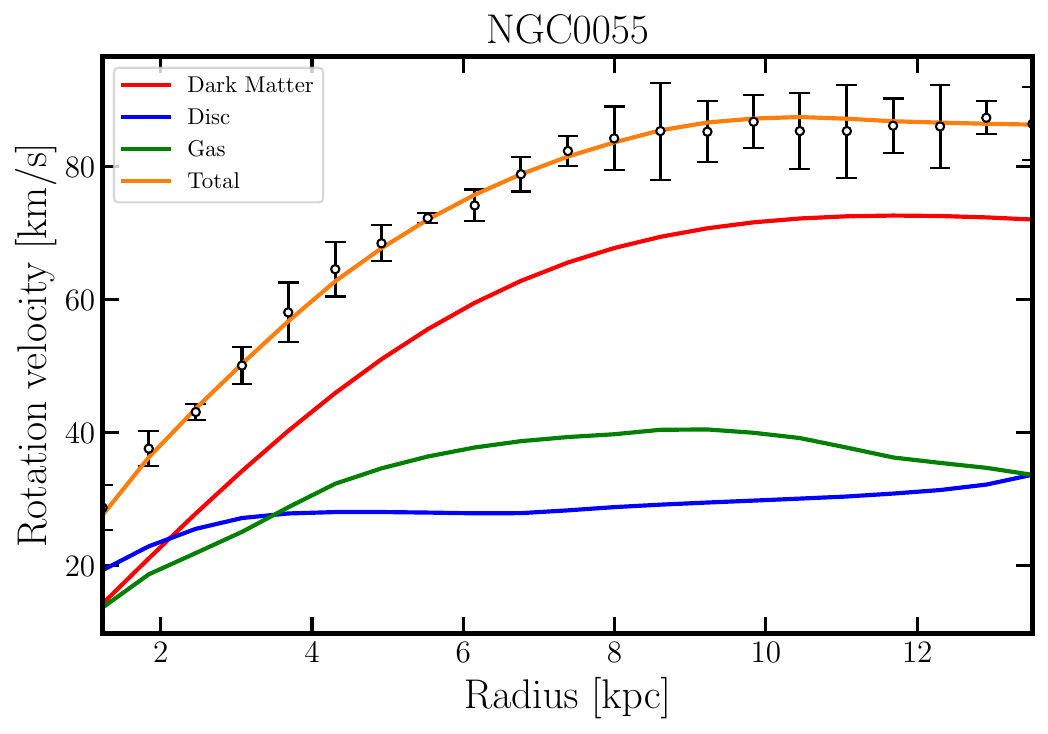}\hfill
    \includegraphics[width=.49\textwidth]{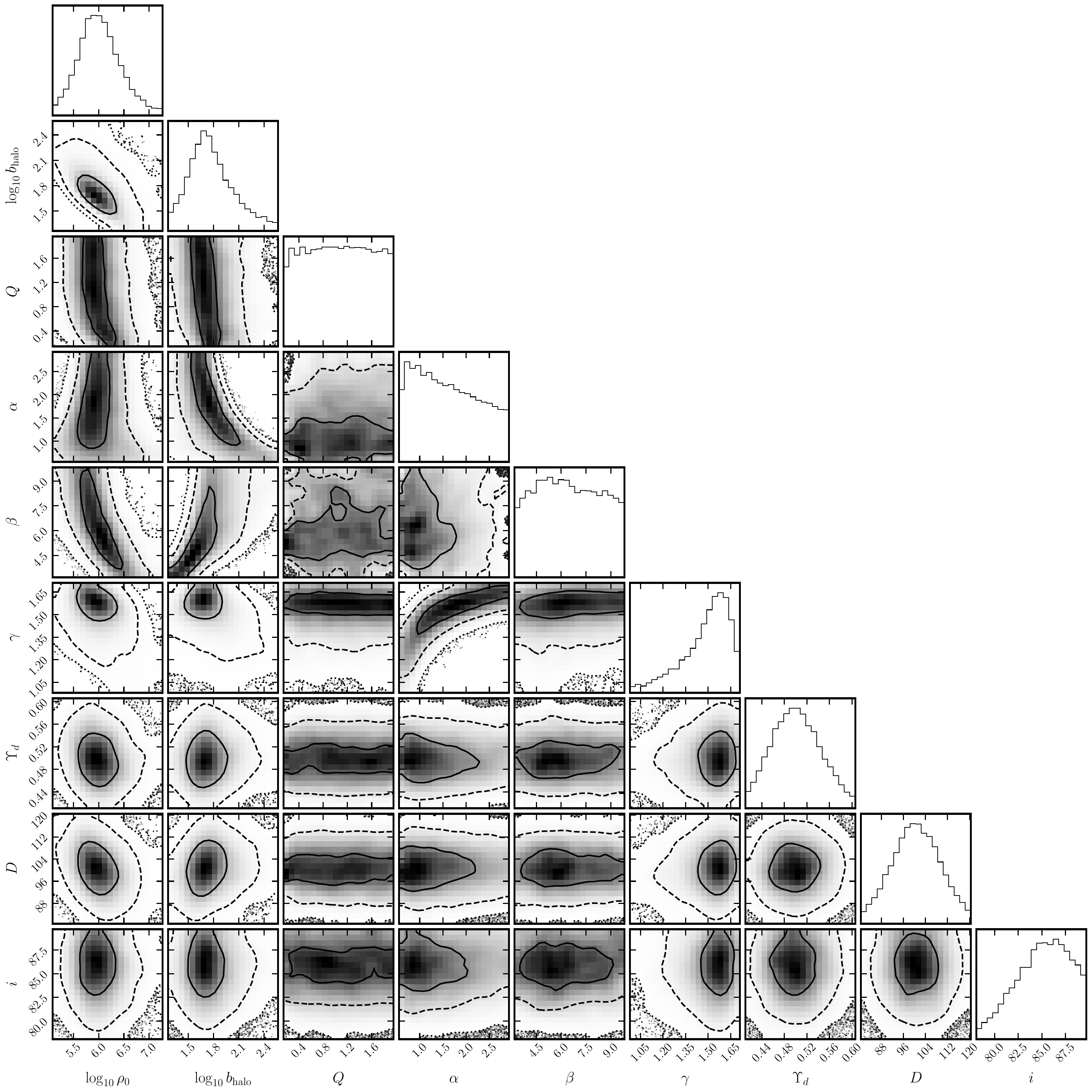}\hfill
    \includegraphics[width=.49\textwidth]{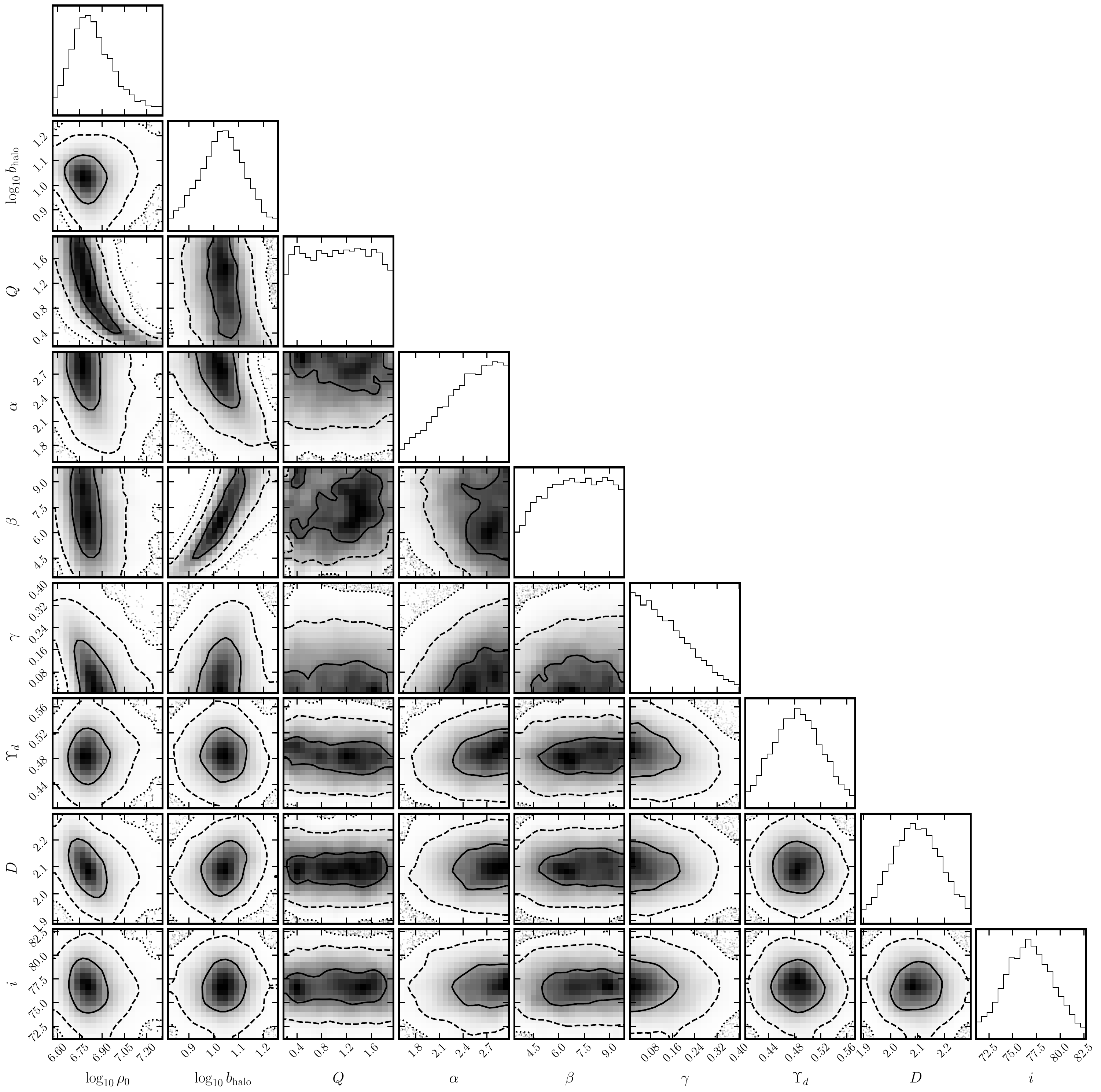}
\caption{Rotation curve fits and \KH{the 1D and 2D marginalized} posterior distributions of the fitting parameters for the UGC12506~(left panels) and NGC0055~(right panels). 
\KH{In the top panels, Green, blue, and red lines denote the contributions of gas, disc, and dark matter respectively. Orange lines represent the total fitted rotation curves.
In the bottom panels, the solid, dashed, and dotted lines denote 68\%, 95\%, and 99\% probabilities, respectively.}
The complete figures of the posterior maps for 115 SPARC galaxies are available in the online journal.
%{Alt text: Rotation curves of two representative galaxies are shown with model fits separating contributions from the stellar disk, gas, and dark matter halo, along with the total predicted curve. Posterior distributions of the dark matter halo parameters are displayed, illustrating that one galaxy is best described by a cuspy halo while the other is better fit by a cored profile.}
} \label{fig:result_exam}
\end{figure*}  
%\footnote{\url{https://drive.google.com/drive/folders/1OxApkk9W-lSxGJq_tBfm4Yrbq128kOmb?usp=sharing}}

\subsection{Total rotation curve}
We perform the fitting analysis for the observed rotation curves by summing the contributions of each stellar and DM component,
\begin{equation}
V^2_\mathrm{tot} = V^2_\mathrm{DM} + \Upsilon_\mathrm{disk}V^2_\mathrm{disk}+ \Upsilon_\mathrm{bul}V^2_\mathrm{bul} + V^2_\mathrm{gas},
\end{equation}
where $V_\mathrm{DM}$ is calculated using Equation~(\ref{eq:v_cric}), while $V_\mathrm{disk},V_\mathrm{bul}$ and $V_\mathrm{gas}$ represent the contributions from disk, bulge, and gas respectively, as tabulated in the SPARC database~\citep{2016AJ....152..157L}.
$\Upsilon_\mathrm{disk}$ and \MM{$\Upsilon_\mathrm{bul}$} are the stellar mass-to-light ratios, which serve as free parameters in this analysis.
Following~\citet{2019MNRAS.482.5106L}, we impose log-normal priors on these ratios, centered around their fiducial values ($\Upsilon_\mathrm{disk}=0.5$ and $\Upsilon_\mathrm{bul}=0.7$) based on \citet{2016PhRvL.117t1101M,2017ApJ...836..152L}.
The standard deviation of these priors is set to 0.1~dex, as suggested by stellar population synthesis models~(e.g., \cite{2001ApJ...550..212B,2014ApJ...788..144M,2019MNRAS.483.1496S}).

Uncertainties in galaxy distance $D$ and disk inclination $i$ affect the radius $R$ and the rotation velocity $V_k$, where $k$ denotes disk, bulge, or gas, respectively.
When the distance $D$ is adjusted to $D^\prime$, $R$ and $V_k$ transform as
\begin{align}
R^\prime = R\frac{D^\prime}{D}; \,\,\, V^\prime_k = V_k \sqrt{\frac{D^\prime}{D}}.
\end{align}
For uncertainties in the disk inclination, with $i$ being adjusted to $i^\prime$, the total observed rotation velocities $V_\mathrm{obs}$ and its observational errors $\delta V_\mathrm{obs}$ transform as
\begin{align}
V_\mathrm{obs}^\prime = V_\mathrm{obs}\frac{\sin(i)}{\sin(i^\prime)}; \,\,\, \delta V^\prime_\mathrm{obs} = \delta V_\mathrm{obs}\frac{\sin(i)}{\sin(i^\prime)}.
\end{align}
Thus, $D$ and $i$ are also free parameters in our fitting procedure. 
We impose Gaussian priors on $D^\prime$ and $i^\prime$ around their mean values \KH{($D$ and $i$) as tabulated in the SPARC database with standard deviations given by their uncertainties.}

In total, our model contains six parameters for the DM halo and four or three parameters for the stellar components, depending on whether the bulge is included or not.

\subsection{Fitting analysis}
Using the standard affine-invariant ensemble sampler in the open source Python package {\it emcee}~\citep{2013PASP..125..306F}, we map the posterior distributions of the free parameters.
\KH{In Bayesian statistics, posterior distributions are determined by the prior and the likelihood function, with the evidence acting as a normalization constant that can be ignored for parameter inference.}
We choose the likelihood function as $\exp(-\frac{1}{2}\chi^2)$, where $\chi^2$ is defined as
\begin{equation}
\chi^2 = \sum_R \frac{[V_\mathrm{obs}(R) - V_\mathrm{tot}(R)]^2}{[\delta V_\mathrm{obs}(R)]^2}.
\end{equation}

\KH{We adopt flat priors for DM halo parameters over the following ranges:
$0 \leq \log_{10}\rho_0/[\mathrm{M_\odot/kpc^{-3}}] \leq 14$\,, $-3\leq \log_{10}b_\mathrm{halo}/[\mathrm{kpc}] \leq 3$, $0.1\leq Q \leq 2.0$, $0.5 \leq \alpha \leq 3.0$, $3.0 \leq \beta \leq 10$, $0 \leq \gamma \leq 2$.}
As described in the previous section, we adopt log-normal priors for $\Upsilon_\mathrm{disk}$ and $\Upsilon_\mathrm{bul}$, and Gaussian priors for $D$ and $i$.

\KH{We performed the MCMC sampling using 280 walkers and 10,000 iterations. After discarding the first 5,000 steps as burn-in, this resulted in approximately $10^6$ posterior samples. We also report basic convergence diagnostics, with an acceptance fraction and visual inspection of the chains.}

\section{Structural properties of dark matter halos}\label{sec:4}

\subsection{Best-fitting models}

Figure \ref{fig:result_exam} shows example fits for SPARC galaxies, UGC12506 and NGC0055.
\KH{Our dynamical analysis indicates that UGC12506 prefers a strongly cusped \MM{DM} density profile ($\gamma = 1.53^{+0.13}_{-0.23}$), whereas NGC0055 clearly favors a cored profile ($\gamma = 0.23^{+0.28}_{-0.16}$).
According to the 1D and 2D marginalized posterior distributions, the DM halo shape parameter ($Q$) and the sharpness and outer slope parameters ($\alpha, \beta$) are broadly distributed within their prior ranges.}
This is because these parameters have a relatively minor impact on the shape of the rotation curve compared to other parameters.
We confirm that even when assuming a spherically symmetric DM halo (i.e., $Q = 1$), our main conclusions and interpretations remain qualitatively unchanged.
The complete figure set of the rotation curves and the posterior distributions for all 115 SPARC galaxies are available in the online journal.

\KH{The best-fit parameters for all 115 SPARC galaxies are summarized in Tables~\ref{tab:Results1}\footnote{\KH{The quoted parameters are estimated using interval estimators derived from the posterior distributions.}}}.
Focusing on the inner DM density slope ($\gamma$), we find a wide range of values across the galaxy sample, spanning from $\gamma = 0.01$ to $1.96$.
This further illustrates the diversity of DM density profiles among low-mass galaxies.
These tables also report the reduced chi-squared values ($\chi^2_\nu = \frac{\chi^2}{N - f}$) for each galaxy, where $N$ is the number of rotation velocity data points and $f$ is the number of free parameters.

%%%%%%%%%%%%%%%%%%%%%%%%%%%%%%%
\begin{figure*}
 \begin{center}
  \includegraphics[scale=0.70]{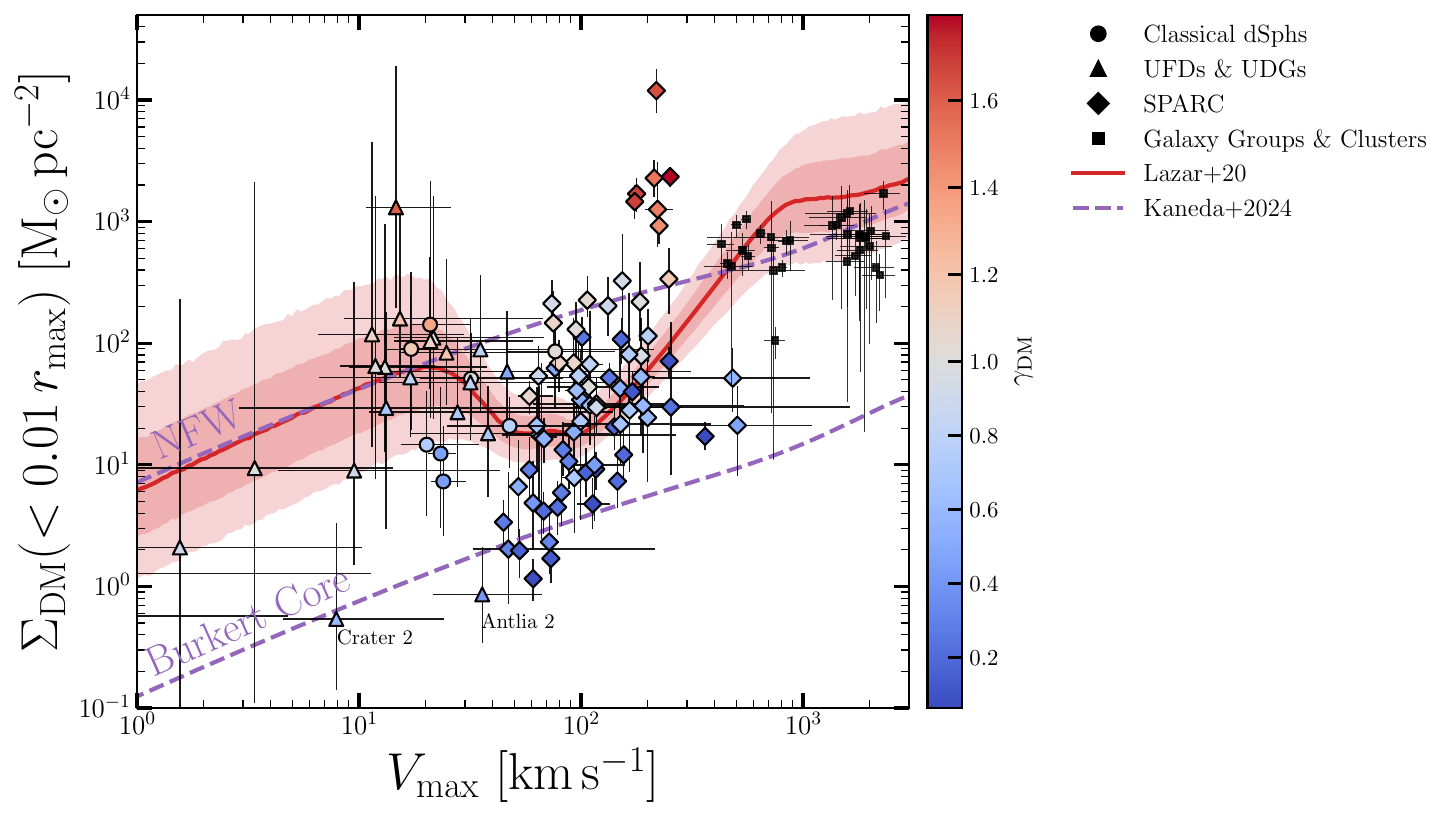} 
 \end{center}
\caption{The central surface density of dark matter halo within 1\% of the radius of the maximum circular velocity, $\Sigma_\textrm{DM}(<0.01r_{V_\textrm{max}})$, as a function of the maximum circular velocity, $V_\textrm{max}$. 
The orange points represent classical dSphs associated with the Milky Way, derived from~\citet{2020ApJ...904...45H}, while the green points correspond to ultra-faint dSphs and ultra-diffuse galaxies, based on \citet{2023ApJ...953..185H}. 
The blue squares indicate the SPARC galaxies analyzed in this work. The black symbols denote galaxy groups and clusters, taken from~\citet{2007ApJ...669..158G}, \citet{2015ApJ...806....4M}, and \citet{2016ApJ...821..116U}.
\MM{The red solid line and the shaded regions are the median and 1, 2, and 3-$\sigma$ halo-to-halo scatter predicted from FIRE-2~\citep{2020MNRAS.497.2393L} hydrodynamical plus dark matter simulations (see text for details). 
In contrast, the two purple dashed lines represent the predictions from \citet{2024PASJ...76.1026K} for the NFW (cuspy) and Burkert (cored) profiles, respectively.}
%{Alt text: Dark matter surface density within 0.01 of the radius where the maximum circular velocity occurs is plotted as a function of the maximum velocity for various galaxy types, including spiral galaxies, classical dwarf spheroidal galaxies, ultra-faint dwarfs, and galaxy clusters. The figure compares observational data with theoretical predictions from NFW and Burkert halo models, as well as hydrodynamical simulations, showing general agreement with cored models for many galaxies.
%-----------------------
}\label{fig:SigmaVmaxRelation}
\end{figure*}
%%%%%%%%%%%%%%%%%%%%%%%%%%%%%%%

\subsection{Central surface density of dark matter halo}
This study aims to identify the cusp-to-core transition phase caused by baryonic feedback, as predicted by $N$-body plus hydrodynamical simulations. To clearly distinguish between cusped and cored dark matter density profiles from observations, \citet{2024PASJ...76.1026K} proposed using the central surface density of the dark matter halo--specifically within 1\% of the radius corresponding to the maximum circular velocity (see also \cite{2015ApJ...803L..11H,2017ApJ...843...97H}) as follows:
\begin{equation}
    \Sigma_\textrm{DM}(<0.01r_{V_\textrm{max}}) = \frac{M_\textrm{DM}(<0.01r_{V_\textrm{max}})}{\pi(0.01r_{V_\textrm{max}})^2},
    \label{eq:sigmaVmax}
\end{equation}
where $r_{V_\textrm{max}}$ is \MM{the radius of maximum circular velocity of the dark matter halo.}
The enclosed mass within $0.01r_{V_\textrm{max}}$ is given by
\begin{equation}
    M_\textrm{DM}(<0.01r_{V_\textrm{max}}) = \int^{0.01r_{V_\textrm{max}}}_{0} 4\pi \rho_\textrm{DM}(r^\prime)r^{\prime2}dr^\prime,
    \label{eq:DMmassSPH}
\end{equation}
where $\rho_\textrm{DM}$ denotes the dark matter density profile (see Equation~\ref{eq:DMH}) assuming spherical symmetry.
\KH{Under the axisymmetric assumption, we generalize this definition by introducing the ellipsoidal radius $r^\prime \equiv m = \sqrt{R^{\prime 2} + z^{\prime 2}/Q^2}$,
where $Q$ is the intrinsic axis ratio of the dark matter halo, and the density is expressed as $\rho_{\rm DM}=\rho_{\rm DM}(m)$ (see Equation~\ref{eq:DMH2}). 
In this case, the enclosed mass within an ellipsoid defined by $m < m_0$ is given by
\begin{equation}
    M_{\rm DM}(<m_0)
    = 4\pi Q \int_{0}^{m_0} \rho_{\rm DM}(m)\, m^2\, dm,
\end{equation}
where we have used the volume element appropriate for similar spheroids.
Throughout this work, we set $m_0 = 0.01\,r_{V_{\max}}$, such that $M_{\rm DM}(<0.01\,r_{V_{\max}})$ refers to the mass enclosed within an ellipsoid whose semi-major axis equals $0.01\,r_{V_{\max}}$.
The quantities $V_{\max}$ and $r_{V_{\max}}$ are determined from the circular velocity defined in Equation~\ref{eq:v_cric}, which is evaluated in the equatorial plane of the axisymmetric potential.}

\KH{Additionally, we also estimate the virial mass $M_\mathrm{DM,200}$ defined as a mass within the ellipsoidal radius $m_{200}$ inside of which the average halo density is 200 times the critical density of the universe.
The virial radius $m_{200}$ is determined by computing both the enclosed dark matter mass and the corresponding enclosed volume assuming an axisymmetric halo geometry. 
Specifically, we evaluate the enclosed mass $M_{\mathrm{DM}}(<m)$ within the ellipsoidal radius $m$, together with the associated spheroidal volume. 
The virial radius $m_{200}$ is then defined as the radius at which the mean enclosed dark matter density equals 200 times the critical density of the Universe. The virial mass $M_\mathrm{DM,200}$ is subsequently obtained as the enclosed dark matter mass within this radius.}

Table~\ref{tab:Results2} summarizes $V_\textrm{max}, r_{V_\textrm{max}}$, $\Sigma_\textrm{DM}(<0.01r_{V_\textrm{max}})$, and \KH{$\log_{10}(M_\ast/M_\textrm{DM,200})$} not only for SPARC galaxies, but also for ultra-faint dwarf dSphs (UFDs), ultra-diffuse galaxies (UDGs), classical dwarf dSphs in the Milky Way, as well as groups and clusters of galaxies.
For the SPARC galaxies, \KH{we estimate these values by marginalizing over the full posterior distributions of the model parameters listed in Table~\ref{tab:Results1}.}
For the Galactic UFDs, UDGs, and classical dSphs, we calculate the corresponding dark matter halo properties using models from  \citet{2020ApJ...904...45H} and \citet{2023ApJ...953..185H}, which are based on stellar kinematics and assume non-spherical dark matter density profiles as described in Equation~(\ref{eq:DMH}). 
In this study, we include 20 UFDs that have more than 10 stellar kinematic data points, ensuring that the number of observational constraints exceeds the number of free model parameters.
On the other hand, following \citet{2024PASJ...76.1026K}, the values for galaxy groups and clusters are derived from the observational \MM{results of}~\citet{2007ApJ...669..158G}, \citet{2015ApJ...806....4M}, and \citet{2016ApJ...821..116U}, assuming spherical symmetry and a cusped NFW dark matter halo, both of which are generally supported by the observational results from X-ray and gravitational lensing analyses.

\section{Cusp-to-core transition}\label{sec:5}
\subsection{$\Sigma_\textrm{DM}(<0.01r_{V_\textrm{max}})$-$V_\textrm{max}$ relation}

Figure~\ref{fig:SigmaVmaxRelation} shows the relation between the dark matter surface density within 1\% of the maximum circular velocity radius, $\Sigma_\textrm{DM}(<0.01r_{V_\textrm{max}})$, and the maximum circular velocity, $V_\textrm{max}$, across a wide mass range spanning from ultra-faint dwarf galaxies (UFDs) to galaxy clusters.
The orange and green points with 1$\sigma$ error bars represent the results for classical dSphs, UFDs, and UDGs in the Milky Way, estimated by~\citet{2020ApJ...904...45H} and~\citet{2023ApJ...953..185H}, respectively.
While the black points correspond to galaxy groups and clusters, the blue points represent the SPARC galaxies analyzed in this study that have reduced $\chi^2$ values less than two.
\KH{We visualize the results using color bars representing different galaxy populations, including ultra-faint dwarfs (UFDs), ultra-diffuse galaxies (UDGs), and SPARC disk galaxies. 
Across these systems, we find that the enclosed dark matter surface density evaluated at small radii, $\Sigma_{\mathrm{DM}}(<0.01\,r_{V_{\mathrm{max}}})$, provides a robust and physically motivated diagnostic of the inner halo structure. In particular, this quantity cleanly separates cuspy and cored dark matter profiles: cuspy halos maintain high central surface densities, while cored systems show a clear suppression of $\Sigma_{\mathrm{DM}}$ at this scale. The consistency of this behavior across a wide range of galaxy masses and morphologies confirms that $\Sigma_{\mathrm{DM}}(<0.01\,r_{V_{\mathrm{max}}})$ is an effective tracer for distinguishing cusp and core formation in dark matter halos.}

The red solid line shows the $\Sigma_\textrm{DM}(<0.01r_{V_\textrm{max}})$–$V_\textrm{max}$ relation derived from a fitting function for the dark matter density profile, based on the results of FIRE-2 zoom-in hydrodynamical simulations~\citep{2020MNRAS.497.2393L}.
Details of the method used to compute the predicted $\Sigma_\textrm{DM}(<0.01r_{V_\textrm{max}})$–$V_\textrm{max}$ relation are provided in Appendix~\ref{sec:App1}.
The shaded regions indicate the 1$\sigma$, 2$\sigma$, and 3$\sigma$ scatters, which are calculated based on the halo-to-halo scatter in the concentration–mass relation.
In addition, the purple dashed lines represent the predicted $\Sigma_\textrm{DM}(<0.01r_{V_\textrm{max}})$–$V_\textrm{max}$ relations for the cusped NFW and cored Burkert profiles, based on the analytic model of the cusp-to-core transition proposed by~\citet{2024PASJ...76.1026K} (see Appendix~\ref{sec:App1} for details).
This model is based on the idea originally proposed by \citet{2014MNRAS.440L..71O} and was later updated with more detailed physical modeling by~\citet{2026arXiv260113868S}.

A notable feature in Figure~\ref{fig:SigmaVmaxRelation} is that a large fraction of the SPARC galaxies lie systematically below the NFW prediction and are instead broadly consistent with the Burkert core profile, despite the substantial uncertainties in both the surface densities and maximum circular velocities across the entire sample.
This behavior suggests that many star-forming galaxies in the SPARC sample possess cored dark matter distributions rather than cuspy ones.
Such a trend is qualitatively consistent with theoretical expectations from the FIRE-2 hydrodynamical simulations, which predict that stellar feedback can transform the central dark matter density structure from a cusp to a core in galaxies with ongoing star formation~\citep{2020MNRAS.497.2393L}.
\textit{Thus, the observational results from SPARC provide important empirical support for the feedback-driven core formation scenario in low- and intermediate-mass galaxies.}

\KH{
In contrast, some SPARC galaxies (e.g., UGC06786, NGC2841, UGC12506, NGC2998, UGC08699, and NGC3521), which exhibit very high central dark matter surface densities,
$\Sigma_{\mathrm{DM}}(<0.01\,r_{V_{\mathrm{max}}}) > 1000~M_\odot\,\mathrm{pc}^{-2}$,
show systematically higher values of $\Sigma_{\mathrm{DM}}(<0.01\,r_{V_{\mathrm{max}}})$ than predicted by CDM simulations that include baryonic physics \citep{2020MNRAS.497.2393L}.
This suggests that the inner dark matter distributions of these galaxies are more centrally concentrated than typically expected from current CDM+baryon models.
The physical origin of these elevated central surface densities is not yet clear; however, several mechanisms may plausibly account for galaxies with such high
$\Sigma_{\mathrm{DM}}(<0.01\,r_{V_{\mathrm{max}}})$.
We return to this point in Section~\ref{sec:5}.}

Most of the ultra-faint dwarf galaxies (UFDs) and galaxy clusters show different behavior.
The UFDs tend to align more closely with the NFW prediction, although the large observational uncertainties prevent definitive conclusions.
This apparent difference may reflect the fact that stellar feedback is less effective at altering the central dark matter distribution in systems with extremely low stellar masses, where the available energy from star formation is insufficient to drive significant cusp-to-core transformations.
Similarly, in massive systems such as galaxy clusters, the deep gravitational potential wells make it difficult for baryonic processes to substantially modify the inner dark matter profiles, resulting in central structures that remain closer to the original cuspy configurations predicted by cold dark matter simulations.
These contrasting trends across different mass scales provide further insight into the mass dependence of core formation efficiency driven by baryonic feedback.

\subsection{Inner Dark Matter Density Slope versus Stellar-to-halo Mass Ratio}

\begin{figure*}
 \begin{center}
  \includegraphics[width=0.9\linewidth]{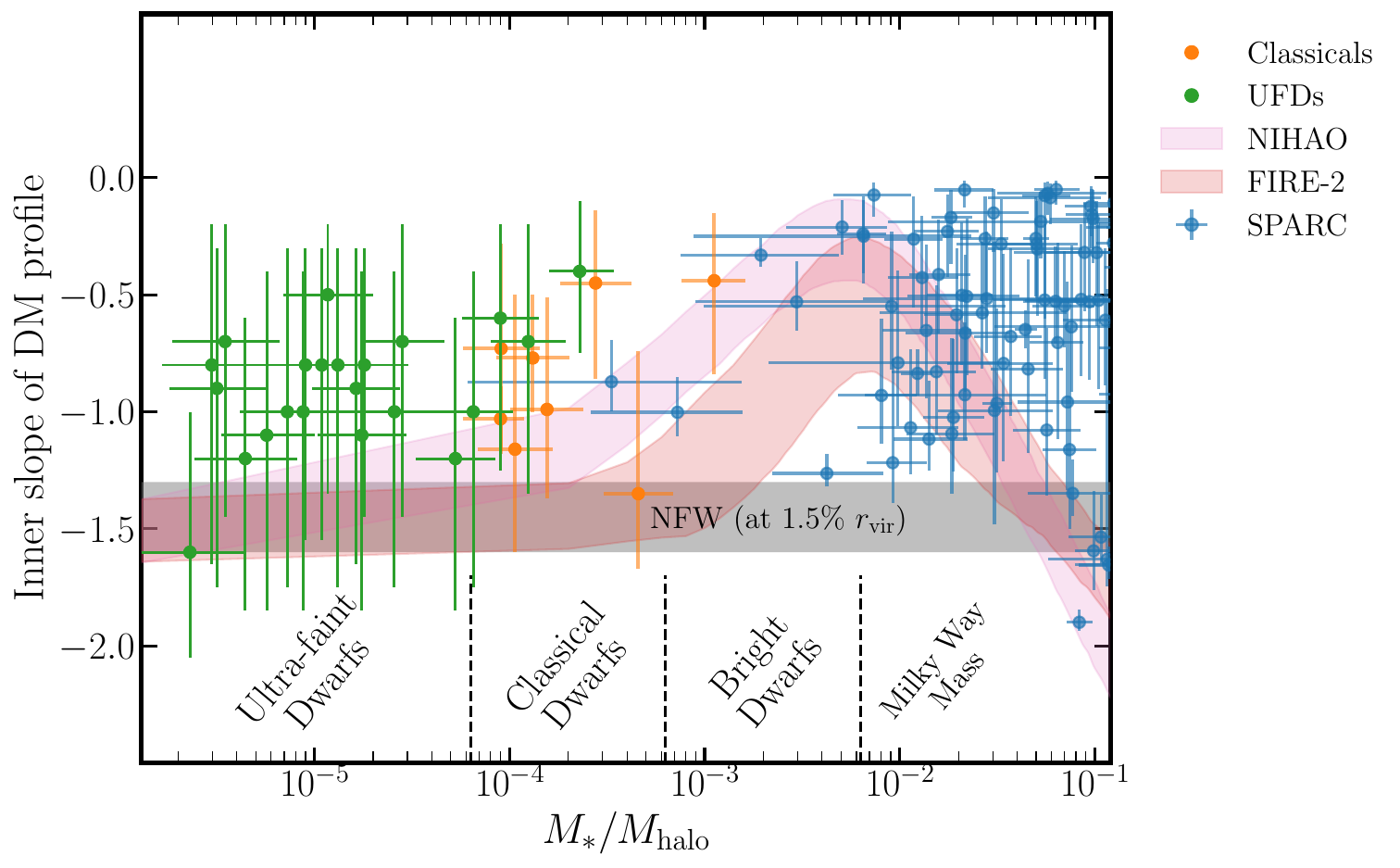} 
 \end{center}
\caption{The inner dark matter density slope at 1.5\% of $r_\textrm{vir}$ is shown as a function of the stellar-to-halo mass ratio.
The filled orange and green circles with error bars are taken from~\citet{2020ApJ...904...45H} and \citet{2023ApJ...953..185H}, respectively, while the filled blue squares represent the results from this work.
The shaded gray band indicates the expected range of inner slopes for NFW profiles, as derived from dark-matter-only simulations~\citep{2016MNRAS.456.3542T}.
The pink and magenta shaded bands show the predicted ranges from the NIHAO~\citep{2016MNRAS.456.3542T} and FIRE-2~\citep{2020MNRAS.497.2393L} simulations, respectively (shown for visual reference).}
%{Alt text: The slope of the inner dark matter density profile is shown as a function of the stellar-to-halo mass ratio. Observational results from different galaxy types are compared with predictions from NFW profiles and results from cosmological simulations such as NIHAO and FIRE-2, highlighting diversity in inner density slopes and departures from pure dark matter predictions.}
\label{fig:gamma_SHMR}
\end{figure*}

Figure~\ref{fig:gamma_SHMR} shows the logarithmic slope of the dark matter density profile as a function of the stellar-to-halo mass ratio ($M_\ast/M_{\rm halo}$) for various types of galaxies.
This figure is based on earlier works by \citet{2020ApJ...904...45H} and \citet{2023ApJ...953..185H}, but is updated here to include SPARC galaxies \MM{analyzed in our study.}
The gray band shows the expected range of dark matter profile slopes for the NFW profiles as derived from 
\MM{dark-matter-only simulations~\citep{2016MNRAS.456.3542T},} while the pink and magenta shaded bands depict the results from NIHAO~\citep{2016MNRAS.456.3542T} and FIRE-2 zoom-in hydrodynamical simulations~\citet{2020MNRAS.497.2393L}.

The orange, green, and blue symbols with $1\sigma$ error bars represent measurements for ultra-faint dwarfs (UFDs), classical dwarf spheroidal galaxies (dSphs), and SPARC galaxies, respectively.
To calculate the stellar mass-halo mass ratios of the UFDs and classical dSphs, we employ the self-consistent abundance matching proposed by \citet{2013MNRAS.428.3121M} and adopt the stellar masses of most dwarf galaxies taken from the literature~(see \cite{2020ApJ...904...45H} and \cite{2023ApJ...953..185H} for details). For several UFDs having no information about stellar masses, we compute their stellar masses based on their luminosities by assuming a stellar mass-to-light ratio of $1.6M_\odot/L_\odot$, which corresponds to the median value for dSphs reported by~\citet{2008MNRAS.390.1453W}.
Stellar masses $M_\ast$ of SPARC galaxies are derived from their 3.6~$\mu$m luminosities, while their dark matter halo masses $M_{\rm halo}$ are inferred from the free parameters $V_{200}$ and $C_{200}$ determined in this study.

In the UFD regime ($M_\ast/M_{\rm halo} \lesssim 10^{-4}$), the predicted inner slope of the dark matter density profile is expected to be largely unaffected by baryonic effects.
However, as shown in this figure, we note that the inner density slopes of all UFDs (Segue~1), except for the one with the lowest $M_\ast/M_{\rm halo}$, are systematically shallower than the predictions from hydrodynamical simulations.

Conversely, while most SPARC galaxies in our sample favour shallower cusped or cored central dark matter density profiles near the Milky Way mass regime, a subset exhibit strongly cusped inner slopes, with some galaxies being widely scattered toward the UFD regime.
This indicates that SPARC galaxies exhibit a wider diversity of dark matter density profiles than predicted by current theoretical simulations.
However, owing to the significant uncertainties in both the inner slopes and the stellar-to-halo mass ratios of the UFDs, it is not possible to draw a robust conclusion regarding the existence of the relation with the currently available data.

%%%%%%%%%%%%%%%%%%%%%%%%%%%%%%%%%%%%%%% 

\section{Discussion}\label{sec:5}

%\end{itemize}

This section discusses possible evidence for the cusp-to-core transition phase and discrepancies between the observations and theoretical predictions, based on the results shown in Figure~\ref{fig:SigmaVmaxRelation}.

\KH{\subsection{Diversity in Central Dark Matter Densities}}

The observations of SPARC galaxies reveal substantial diversity in their central dark matter densities and inner density slopes, with some galaxies exhibiting shallow cores and others retaining relatively steep cusps. 
\KH{In the context of CDM, repeated episodes of baryonic feedback from supernovae or stellar winds can induce fluctuations in the gravitational potential, leading to a redistribution of dark matter and a reduction of the central density, effectively transforming cuspy profiles into cored ones \citep{2016MNRAS.459.2573R,2018MNRAS.473.4392S,2020MNRAS.495...58S}.
On the other hand, baryonic processes can also act in the opposite direction: adiabatic contraction associated with the buildup of stellar and gaseous mass in the inner regions of disk galaxies (e.g., \cite{2004ApJ...616...16G,2015MNRAS.454.2981C}).
In this scenario, adiabatic contraction enhances the central dark matter density, resulting in a steeper inner density profile.
Therefore, while baryonic effects provide plausible mechanisms for modifying central dark matter densities.
However, they do not straightforwardly account for the full observed diversity, as most hydrodynamical simulations predict a relatively narrow range of inner density structures at fixed halo mass.}

Although baryonic processes may help resolve the cusp to core problem, their connection to the observed diversity remains uncertain and warrants further investigation. It is not the expulsion of gas itself, but rather the oscillatory motion of baryons, driven by repeated episodes of star formation and feedback, that plays the central role. This mechanism was proposed by \citet{2010Natur.463..203G} and \citet{2012MNRAS.421.3464P}, who argued that time dependent baryonic potentials could induce the transition from a central cusp to a core. In a separate line of work, \citet{2014ApJ...793...46O} developed a physically grounded model in which Landau resonance serves as the key mechanism for energy transfer between baryonic oscillations and the dark matter halo. These coherent fluctuations in the gravitational potential can drive dark matter particles outward, resulting in a flattened central density profile. This interaction between waves and particles leads to a predictive relationship between the oscillation period and the size of the resulting core. 

Although this scenario is supported by many recent simulations that reproduce oscillatory star formation histories consistent with the conditions required for this mechanism, a critical question remains unresolved: how efficiently is the energy released by supernova explosions converted into gravitational potential fluctuations sufficient to drive the transformation from cusp to core? Investigating the connection between this efficiency and the observed diversity in the central dark matter structures of dwarf galaxies would provide valuable insights into the physical origin of the diversity problem. The energy conversion efficiency in this context has not yet been fully quantified. This issue is examined in detail by Shinozaki et al.\ (in prep.), who focus on quantifying the energy conversion efficiency relevant to feedback-driven core formation.

SIDM models offer a possible explanation for the diversity observed in SPARC galaxies, as they naturally predict a range of core sizes and inner density slopes that depend on the baryonic mass of the galaxy (e.g., \cite{2017PhRvL.119k1102K,2019PhRvX...9c1020R}), or, in regimes with sufficiently large self-interaction cross sections, through gravothermal core collapse \citep{2025PhRvD.111j3041R}. However, it remains unclear whether SIDM alone can fully account for the observed scatter, or whether a combination of baryonic processes and dark-sector physics is required.

As already noted in previous studies (e.g., \cite{2019MNRAS.490..231K,2019MNRAS.484.1401R,2020ApJ...904...45H,2023ApJ...953..185H}), the Milky Way dwarf satellites, including both classical dSphs and ultra-faint dwarfs, also exhibit significant diversity in their central dark matter densities and inner slopes, despite the associated observational uncertainties. This observed diversity in the inner dark matter structures of Milky Way satellites, as well as field dwarfs, can in principle be explained within both the standard CDM framework with baryonic feedback and alternative dark matter models such as SIDM~(e.g., \cite{2020PhRvD.101f3009N,2021MNRAS.503..920C,2023ApJ...949...67Y}). 
However, distinguishing between these scenarios remains challenging with current observational data, and no definitive conclusion has yet been reached regarding the dominant mechanism responsible for the diversity.

Some ultra-diffuse dwarf galaxies, such as Crater~2 and Antlia~2, exhibit properties indicative of strong tidal disruption. These systems possess unusually large half-light radii and low surface brightnesses, suggesting that they have undergone significant mass loss due to tidal stripping by the Milky Way’s gravitational potential. Their present-day structural and kinematic properties may therefore no longer reflect the initial conditions of isolated halo evolution assumed in simulations like those of \citet{2020MNRAS.497.2393L}. The effects of tidal heating, mass loss, and phase-space mixing must be carefully accounted for before directly comparing such disrupted systems to predictions based on equilibrium models.

Finally, in the regime of galaxy groups and galaxy clusters, most systems are found to be consistent with cusped NFW dark matter halos.
This consistency implies that such massive systems are less susceptible to baryonic feedback due to their deep gravitational potential wells predominantly governed by dark matter.
In Figure~\ref{fig:SigmaVmaxRelation}, although the model by \citet{2020MNRAS.497.2393L} deviates slightly from the NFW prediction of \citet{2024PASJ...76.1026K}\footnote{\KH{The analysis in this paper was limited to systems with maximum circular velocities of $V_{\mathrm{max}}\lesssim 1500$ km~s$^{-1}$}}, it should be noted that the former primarily targets galactic and sub-galactic mass scales rather than groups or clusters.
Therefore, the red solid line and the shaded region shown in Figure~\ref{fig:SigmaVmaxRelation} represent extrapolated predictions for systems with $V_\mathrm{max} \gtrsim 500$ km~s$^{-1}$ and should be interpreted with caution.
This trend reinforces the notion that the cusp-to-core transition is a mass-dependent phenomenon, being most prominent at galaxy mass scales, and increasingly suppressed at the high-mass end where baryonic processes are less effective in reshaping dark matter distributions.

These considerations highlight the importance of accounting for mass scale, tidal effects, and observational limitations when interpreting the inner dark matter density structures of galaxies. 
Our findings suggest that baryonic feedback likely plays a significant role in producing the observed diversity, but a complete understanding will require improved simulations that incorporate both baryonic and dark sector physics, along with deeper and more precise observational constraints, particularly for low-mass and environmentally affected systems.

\subsection{Limitations and Caveats in Theory and Observations}

It should be noted that the simulations by \citet{2020MNRAS.497.2393L} modeled a total of 54 isolated galaxies, with typical halo masses in the range $M_{\rm halo} \sim 10^{4.5}$–$10^{11}\,M_\odot$. 
These simulations were not intended to model LSB galaxies or more massive systems with halo masses typically exceeding $10^{11}\,M_\odot$. 
Given these differences in mass scales and the associated baryonic physics, direct comparisons between the simulated dwarf galaxies and the observed SPARC galaxies must be approached with caution. 
In particular, mechanisms such as feedback-driven core formation may operate differently or less efficiently in more massive halos, and the dynamical timescales that govern their central structures can also differ. 

The recent results from the FIRE-3 simulations~\citep{2024arXiv240416247S} have emphasized that some rotation curves, particularly in low-mass galaxies, may not be reliable tracers of the underlying dark matter distribution. These studies found that in galaxies rich in HI gas with well-ordered disks, the deviation from the true rotation curve is typically around 10\,\%, whereas in galaxies affected by dynamical disequilibrium, non-circular motions, or strong magnetic fields, the deviation can exceed 50\,\%. In particular, they demonstrated that the galaxies that sparked the so-called diversity problem (e.g., IC~2574, UGC~5721, and UGC~8837) can be explained by these effects. According to these recent simulation results, whether the diversity problem genuinely exists remains a subject of ongoing debate.

\section{Summary and Conclusion}\label{sec:6}

In this study, we have investigated the dark matter density distributions in a sample of late-type galaxies using high-quality rotation curves from the SPARC database. By adopting a flexible dark matter halo model that allows for variations in inner slopes and axis ratios, and \MM{by} performing detailed Bayesian fits to the observed kinematics, we have characterized the inner dark matter structures across a wide range of galaxy masses.

Our analysis reveals a large diversity in the inner dark matter density slopes among the SPARC galaxies, ranging from steep cusps to shallow cores. 
We find that a substantial fraction of these galaxies favor cored or shallower cusped profiles. 
When applying the central surface density of the dark matter halo, $\Sigma_\mathrm{DM}(<0.01r_{V_\mathrm{max}})$, to the results from the SPARC galaxies, as well as to the estimated dark matter halos of the Galactic dwarf satellites and galaxy groups and clusters, we find a notable feature that a large fraction of the SPARC galaxies lie systematically below the NFW prediction and are broadly consistent with the Burkert core profile. 
On the other hand, systems at both lower and higher mass scales than the SPARC galaxy mass range are roughly consistent with the NFW cusped predictions.

These observational results provide important empirical support for the baryonic feedback-driven core formation scenario predicted by $\Lambda$CDM plus hydrodynamical simulations, despite the substantial uncertainties in both the surface densities and maximum circular velocities across the sample. 
Our findings suggest that baryonic processes may play a significant role in shaping the central dark matter structures and could account for much of the observed diversity, although some discrepancies still remain.
We also emphasize that limitations exist in interpreting rotation curves as direct tracers of the underlying gravitational potential, particularly in low-mass systems affected by non-circular motions and pressure support. 

A comprehensive understanding of the origin and diversity of dark matter density profiles will require improved theoretical modeling—including both baryonic and dark sector physics—as well as deeper and more precise observational constraints. 
Future spectroscopic observations of dwarf satellites with facilities such as Subaru-PFS~(\cite{2014PASJ...66R...1T,2016SPIE.9908E..1MT}) and 30-m class telescopes~(\cite{2016SPIE.9908E..1VS}) will further tighten constraints on their dark matter density profiles. In addition, high-precision astrometric and imaging data from missions such as the Roman Space Telescope~(\cite{2024SPIE13092E..0SS}) will provide complementary constraints on the dynamics and structure of these systems~(\cite{2019JATIS...5d4005W}). Furthermore, next-generation radio telescopes such as the Square Kilometre Array~(\cite{2019arXiv191212699B}) will significantly improve the precision of rotation curve measurements for late-type galaxies~(\cite{2018MNRAS.473.3256O}), enabling more robust tests of core formation scenarios.

%%%%%%%%%%%%%%%%%%%%%%%%%%%%%%%%%%%%%%% 

\section*{Supplementary data} 

The following supplementary data is available at PASJ online.

Figures E1-E230.

\begin{ack}
We are grateful to the referee for the careful reading of our
paper and thoughtful comments.
We would like to give special thanks to Shigeki
Matsumoto and Masahiro Takada for useful discussions.
This work is supported by Grant-in-Aid for Scientific Research from the Ministry of Education, Culture, Sports, Science, and Technology (MEXT), Japan, grant numbers 20H01895, 21K13909, 23H04009, JP24K00669, JP25H01553~(K.H.), 23KJ0280~(Y.K.), and JP24K07085~(M.M.).

\end{ack}

%\section*{Funding}
% This research was supported by ...

%\section*{Data availability} 
% The data underlying this article are available ...  
% Sample Data Availability Statements 
% https://academic.oup.com/pages/open-research/research-data#Data%20Availability%20Statements
%%%% 

% Any journal's BST file (e.g., apj.bst) can be used as PASJ's BST is unavailable.    
% \bibliographystyle{****}
% \bibliography{****}

%\clearpage
\appendix %%%%%%%%%%%%%%%%%%%%%%%%%%%%%%%%%%%%%%%%%%%%%%%%%%%%%%%%
\section{The theoretical $\Sigma_\textrm{DM}(<0.01r_{V_\textrm{max}})$-$V_\textrm{max}$ relation} \label{sec:App1}
In order to compute $\Sigma_\textrm{DM}(<0.01r_{V_\textrm{max}})$-$V_\textrm{max}$ relation from theoretical predictions, we adopt two representative frameworks: one based on high-resolution $N$-body plus hydrodynamical simulations, and the other based on analytic models of the cusp-to-core transition.
We here introduce these two frameworks in details.

\subsection{Based on Lazar et al. (2020)}
Based on high-resolution galaxy simulations from the FIRE-2 project~\citep{2018MNRAS.480..800H}, \citet{2020MNRAS.497.2393L} introduced an analytic dark matter density profile, named as {\it core-Einasto} profile, to model the diverse inner structures of $\Lambda$CDM haloes, especially those influenced by baryonic feedback. 
This profile generalizes the classic Einasto form~\citep{1965TrAlm...5...87E} by incorporating a core radius parameter, $r_c$, which allows for a pronounced constant density core in the resolved innermost radius.
The core-Einasto profile is given by
\begin{equation}
\rho(r) = \rho_s \exp\left\{ -\frac{2}{\alpha_E} \left[ \left( \frac{r + r_c}{r_s} \right)^{\alpha_E} - 1 \right] \right\},
\label{eq:coredEinasto}
\end{equation}
where $\rho_s, r_s, r_c$, and $\alpha_E$ are a scale density, scale radius, core radius, and shape parameter, respectively.
In this work, we adopt a fixed value of $\alpha_E = 0.16$, following \citet{2020MNRAS.497.2393L}.

Regarding the parameters $r_c$ and $r_s$, \citet{2020MNRAS.497.2393L} provide fitting functions for them as a function of the stellar-to-halo mass ratio, $M_\ast/M_\mathrm{halo}$, to account for the influence of baryonic feedback on the inner dark matter density distribution.
For $r_c$, the fitting function is given by
\begin{equation}
    r_c(x) = 10^{1.21} \left(0.71 + \frac{x}{7.2\times10^{-3}}\right)^{-2.31} \left(\frac{x}{0.011}\right)^{1.55} \,\,\, \textrm{[kpc]},
    \label{eq:coreradius}
\end{equation}
where $x=M_\ast/M_\mathrm{halo}$~(see Equation~(12) and Table~1 in \citet{2020MNRAS.497.2393L} for details).
On the other hand, the fitting function for $r_s$ is expressed in terms of the ratio between $r_s$ and $r_{-2}$, where $r_{-2}$ is defined as the radius at which the log-slope of the Einasto dark matter density is equal to $-2$:
\begin{equation}
%    [r_s/r_{-2}](x) 
    \left[\frac{r_s}{r_{-2}}\right](x)= \left(1+\frac{x}{0.044}\right)^{-31.79} + 1.51\left(\frac{x}{0.28}\right)^{0.40},
    \label{eq:scaleradius}
\end{equation}
where $x=M_\ast/M_\mathrm{halo}$~(see Equation~(13) and Table~2 in \citet{2020MNRAS.497.2393L} for details).

They argued that stellar feedback in dark matter haloes affects the halo concentration through the gravitational coupling of dark matter to the rapidly evolving central gravitational potential.
To quantify this, they adopted the halo concentration parameter defined as $c_{200} \coloneqq r_{200} / r_{-2}$, where $r_{200}$ is the radius enclosing a mean density 200 times the critical density of the universe. This definition was applied to both the FIRE-2 hydrodynamical simulation ($c_\mathrm{F2}$) and their corresponding dark-matter-only counterparts ($c_\mathrm{DM}$).
Using these parameters, they also provide the fitting function of $c_\mathrm{F2}/c_\mathrm{DM}$ as a function of $M_\ast/M_\mathrm{halo}$:
\begin{equation}
%[c_\mathrm{F2}/c_\mathrm{DM}](x) = 
\left[\frac{c_\mathrm{F2}}{c_\mathrm{DM}}\right](x)=\left(1+\frac{x}{4.28\times10^{-3}}\right)^{-1.80} + 0.374\left(\frac{x}{4.28\times10^{-3}}\right)^{0.66}, 
\label{eq:CcoredEinasto}
\end{equation}
where $x=M_\ast/M_\mathrm{halo}$~(see Equation~(14) and Table~3 in \citet{2020MNRAS.497.2393L} for details).
In this work, we assume that $c_\mathrm{DM}$ follows the concentration-mass relation from \citet{2020Natur.585...39W}, which extends over twenty orders of magnitude in dark matter halo mass.
The relation is expressed as a sixth-order polynomial in the logarithm of the halo mass:
\begin{equation}
    c_\mathrm{DM}(M_{200}) = \sum^{5}_{i=0}c_i\left[\ln\frac{M_{200}}{h^{-1}M_\odot}\right]^i,
    \label{eq:cMrelation}
\end{equation}
where $M_{200}$ is the total mass within $r_{200}$, and the coefficients are given by $c_i=[27.112,-0.381,-1.853\times10^{-3},-4.141\times10^{-4},-4.334\times10^{-6},3.208\times10^{-7}]$.

Furthermore, to compute the stellar-to-halo mass ratio across the full range of halo masses considered in this work, we adopt the self-consistent abundance matching model proposed by \citet{2013MNRAS.428.3121M}:
\begin{equation}
    \frac{M_\ast}{M_\mathrm{200}} = 2\times0.0351\left[\left(\frac{M_\mathrm{200}}{11.590}\right)^{-1.376} + \left(\frac{M_\mathrm{200}}{11.590}\right)^{0.608}\right]^{-1},
    \label{eq:SHMR}
\end{equation}
which is calibrated at the redshift $z = 0$.

In summary, given a halo mass $M_{200}$, the concentration $c_\mathrm{DM}$ and the stellar-to-halo mass ratio $M_\ast/M_\mathrm{halo}$ can be estimated using Equations~(\ref{eq:cMrelation}) and (\ref{eq:SHMR}), respectively.
The virial radius $r_{200}$ can also be computed via $r_{200}=(GM_{200}/100H_0^2)^{1/3}$, where $G$ is the gravitational constant and $H_0$ is the Hubble constant.
Utilizing Equation~(\ref{eq:coreradius}), (\ref{eq:scaleradius}), and (\ref{eq:CcoredEinasto}), we can uniquely determine the dark matter halo parameters $(\rho_s, r_s, r_c)$ that define the core-Einasto profile in Equation~(\ref{eq:coredEinasto}).
Once these parameters are specified, we can compute key dynamical and structural quantities such as the maximum circular velocity $V_\mathrm{max}$ and the inner dark matter surface density $\Sigma_\mathrm{DM}(<0.01\,r_{V_\mathrm{max}})$.

\subsection{Based on Kaneda et al. (2024)}
\label{subsec:CCtransmodel}
%In this section, we construct a scaling relation for cored haloes based on the $c$--$M$ relation.
%In previous sections, we compared $\Lambda$CDM model predictions with observations assuming the dark matter density profile to have a central cusp.
%However, observations of disc galaxies, dwarf galaxies, and low surface brightness disc galaxies generally indicate the existence of a core in the centre of the dark matter haloes (see section \ref{sec:intro}).
%Most scaling relations hold for the cored density distribution. 
%While several solutions to solve the cusp--core problem are introduced in section \ref{sec:intro}, we especially focus on the baryonic solution in this section.
%We stand on the hypothesis that dark matter haloes formed primarily with a cuspy density distribution and then some dynamical processes lead to the formation of a core at the centre of the dark matter haloes.
%We simplify the complicated physical processes and provide a model to convert the parameters of a cuspy profile, namely, the scale density and the scale radius, to the parameters of a core profile, the central density and the core radius.

%\subsection{Cusp--to--core transition model} 
In this section, the cusp--to--core transition model~\citep{2024PASJ...76.1026K} is described.
They adopt the NFW profile as an initial state and the Burkert profile as a final state.
The parameters of initial NFW profile, namely, the scale density $\rho_\mathrm{N}$ and the scale radius $r_\mathrm{N}$, are determined by a concentration-mass relation.
They assume that these parameters can be converted into those of the Burkert profile through the following equations: $r_\mathrm{B} = \eta r_\mathrm{N}$ and $\rho_\mathrm{B} = \zeta \rho_\mathrm{N}$, where $r_\mathrm{B}$ and $ \rho_\mathrm{B}$ are the scale radius and density of the Burkert profile.
They impose two physically motivated conditions for the cusp-to-core transition, assuming that the gas fraction in the dark matter halo is sufficiently small.
First, only the central density distribution is changed, while the outer regions remain unchanged.
Second, the virial mass of the dark matter halo, $M_{200}$, is conserved throughout the transition.
From the first condition, equating the density distribution of the NFW profile and the Burkert profile at $r=r_{200}$, we have
\begin{equation}
    \frac{\rho_\mathrm{N}}{\frac{r_{200}}{r_\mathrm{N}}\left(\frac{r_{200}}{r_\mathrm{N}}+1\right)^2} = \frac{\rho_\mathrm{B}}{\left(\frac{r_{200}}{r_\mathrm{B}}+1\right)\left(\frac{r_{200}^2}{{r_\mathrm{B}}^2}+1\right)}  \label{eq:(i)}
\end{equation}
As a requirement that meets the second condition, they adopt the conservation of virial masses, in other words, they equate enclosed mass for the NFW profile:
\begin{equation}
    M(<r) = 4\pi\rho_\mathrm{N} r_\mathrm{N}^3 f_\mathrm{N}\left(\frac{r}{r_\mathrm{N}}\right) \label{eq:m_nfw},
\end{equation}
and the enclosed mass for the Burkert profile:
\begin{equation}
    M(<r) = 4 \pi \rho_\mathrm{B} r_\mathrm{B}^{3} f_\mathrm{B}\left(\frac{r}{r_\mathrm{B}}\right) \label{eq:m_burkert},
\end{equation}
at $r=r_{200}$ and yield
\begin{equation}
    \rho_\mathrm{N} r_\mathrm{N}^3 f_\mathrm{N}\left(\frac{r_{200}}{r_\mathrm{N}}\right) = \rho_\mathrm{B} r_\mathrm{B}^{3} f_\mathrm{B}\left(\frac{r_{200}}{r_\mathrm{B}}\right). \label{eq:(ii)}
\end{equation}
Eliminating $\rho_\mathrm{B}$ from equations (\ref{eq:(i)}) and (\ref{eq:(ii)}) and substituting $r_\mathrm{B} = \eta r_\mathrm{N}$, we obtain
\begin{equation}
    g(\eta) - \frac{ f_\mathrm{N}(c_{200})}{ f_\mathrm{B}(c_{200}/\eta)} = 0,  \label{eq:bisection_method}
\end{equation}
where 
\begin{equation}
    g(\eta) \equiv \frac{(c_{200}+\eta)(c_{200}^2+\eta^2)}{c_{200}(c_{200}+1)^2}.
\end{equation}

They solve equation (\ref{eq:bisection_method}) for $\eta$ numerically at each $c_{200}$ value derived by a c-M relation.
Here, note that $r_{200}$ is determined from a given $M_{200}$ using 
\begin{equation}
     M_{200}\equiv\frac{4}{3}\pi200\rho_{\mathrm{crit,0}}r_{200}^3. \label{eq:m200}
\end{equation}
Then substituting $r_\mathrm{N}$, $r_\mathrm{B}$, and $r_{200}$ into equation (\ref{eq:(i)}) or equation (\ref{eq:(ii)}), we obtain $\rho_\mathrm{B}$.

In this way, one can derive the value of $\rho_\mathrm{B}$ and $r_\mathrm{B}$ from a given combination of $\rho_\mathrm{N}$ and $r_\mathrm{N}$.
In this paper, we use the c-M relation taken from \citet{2023MNRAS.518..157M} to get a combination of $\rho_\mathrm{N}$ and $r_\mathrm{N}$.
%%%% 

\begin{comment}
\begin{figure*}[t!]
 \begin{center}
  \includegraphics[scale=0.7]{./figures/RotCurve_sum_01.pdf} 
 \end{center}
\caption{ The best rotation-curve fits of SPARC galaxies.}
\label{fig:rot01}
\end{figure*}

\begin{figure*}[t!]
 \begin{center}
  \includegraphics[scale=0.7]{./figures/RotCurve_sum_02.pdf} 
 \end{center}
\caption{ The same as Fig.~\ref{fig:rot01}, but the other SPARC samples.
}\label{fig:rot02}
\end{figure*}

\begin{figure*}[t!]
 \begin{center}
  \includegraphics[scale=0.7]{./figures/RotCurve_sum_03.pdf} 
 \end{center}
\caption{ The same as Fig.~\ref{fig:rot01}, but the other SPARC samples.
}\label{fig:rot03}
\end{figure*}

\begin{figure*}[t!]
 \begin{center}
  \includegraphics[scale=0.7]{./figures/RotCurve_sum_04.pdf} 
 \end{center}
\caption{ The same as Fig.~\ref{fig:rot01}, but the other SPARC samples.
}\label{fig:rot04}
\end{figure*}

\begin{figure*}[t!]
 \begin{center}
  \includegraphics[scale=0.7]{./figures/RotCurve_sum_05.pdf} 
 \end{center}
\caption{ The same as Fig.~\ref{fig:rot01}, but the other SPARC samples.
}\label{fig:rot05}
\end{figure*}

\begin{figure*}[t!]
 \begin{center}
  \includegraphics[scale=0.7]{./figures/RotCurve_sum_06.pdf} 
 \end{center}
\caption{ The same as Fig.~\ref{fig:rot01}, but the other SPARC samples. 
}\label{fig:rot06}
\end{figure*}

\begin{figure*}[t!]
 \begin{center}
  \includegraphics[scale=0.7]{./figures/RotCurve_sum_07.pdf} 
 \end{center}
\caption{ The same as Fig.~\ref{fig:rot01}, but the other SPARC samples.}
\label{fig:rot07}
\end{figure*}
\end{comment}

% Table of the best-fit parameters
%\tiny
\fontsize{6pt}{8pt}\selectfont
\begin{longtable}{*{12}{c}}
\caption{The best-fit parameters for all 115 SPARC galaxies.}
\label{tab:Results1}
\hline
Galaxy Name & $\log_{10}\,\rho_0$ & $\log_{10}\,b_\mathrm{halo}$ & $Q$ & $\alpha$ & $\beta$ & $\gamma$ & $\Upsilon_\mathrm{disk}$ & $\Upsilon_\mathrm{bulge}$ & $D_\odot$ & Inclination & $\chi^2_\nu$  \\ 
& [$M_\odot$~kpc$^{-3}$] & [kpc]     &   &    &  &  & [$M_\odot/L_\odot$] & [$M_\odot/L_\odot$] & [Mpc] & [deg] & \\ 
\hline
\endfirsthead
\hline
Galaxy Name & $\log_{10}\,\rho_0$ & $\log_{10}\,b_\mathrm{halo}$ & $Q$ & $\alpha$ & $\beta$ & $\gamma$ & $\Upsilon_\mathrm{disk}$ & $\Upsilon_\mathrm{bulge}$ & $D_\odot$ & Inclination & $\chi^2_\nu$  \\ 
& [km~s$^{-1}$] &     &   &    &  &  & [$M_\odot/L_\odot$] & [$M_\odot/L_\odot$] & [Mpc] & [deg] & \\ 
\hline
\endhead
\hline
\endfoot
\hline
\endlastfoot

D631-7 & $6.88^{+0.16}_{-0.11}$ & $0.94^{+0.13}_{-0.10}$ & $1.10^{+0.59}_{-0.59}$ & $2.49^{+0.36}_{-0.58}$ & $6.83^{+2.11}_{-2.36}$ & $0.07^{+0.09}_{-0.05}$ & $0.47^{+0.04}_{-0.04}$ & $\cdots$ & $7.67^{+0.17}_{-0.17}$ & $56.48^{+2.75}_{-2.64}$ & 0.85\\ 
DDO064 & $6.47^{+0.83}_{-0.93}$ & $1.60^{+0.93}_{-0.85}$ & $1.07^{+0.62}_{-0.63}$ & $1.84^{+0.76}_{-0.84}$ & $6.31^{+2.39}_{-2.29}$ & $0.62^{+0.24}_{-0.31}$ & $0.50^{+0.05}_{-0.05}$ & $\cdots$ & $7.45^{+1.93}_{-1.94}$ & $60.59^{+4.63}_{-4.77}$ & 0.64\\ 
DDO154 & $7.08^{+0.20}_{-0.18}$ & $0.75^{+0.12}_{-0.14}$ & $1.09^{+0.59}_{-0.57}$ & $2.00^{+0.53}_{-0.42}$ & $6.92^{+1.98}_{-2.12}$ & $0.27^{+0.17}_{-0.17}$ & $0.49^{+0.05}_{-0.04}$ & $\cdots$ & $4.04^{+0.18}_{-0.18}$ & $63.96^{+2.64}_{-2.71}$ & 1.67\\ 
DDO161 & $6.44^{+0.36}_{-0.25}$ & $1.15^{+0.19}_{-0.18}$ & $1.10^{+0.60}_{-0.63}$ & $2.17^{+0.53}_{-0.67}$ & $6.62^{+2.25}_{-2.35}$ & $0.42^{+0.22}_{-0.24}$ & $0.50^{+0.05}_{-0.05}$ & $\cdots$ & $7.18^{+1.64}_{-1.83}$ & $69.02^{+9.06}_{-8.77}$ & 0.21\\ 
DDO168 & $7.45^{+0.22}_{-0.15}$ & $0.63^{+0.07}_{-0.07}$ & $0.90^{+0.68}_{-0.54}$ & $2.77^{+0.17}_{-0.31}$ & $7.71^{+1.59}_{-2.22}$ & $0.05^{+0.07}_{-0.04}$ & $0.49^{+0.05}_{-0.04}$ & $\cdots$ & $4.15^{+0.19}_{-0.20}$ & $55.28^{+5.37}_{-5.18}$ & 15.69\\ 
ESO079-G014 & $7.45^{+0.34}_{-0.31}$ & $1.03^{+0.19}_{-0.20}$ & $1.17^{+0.58}_{-0.67}$ & $2.18^{+0.52}_{-0.51}$ & $5.75^{+2.67}_{-1.93}$ & $0.20^{+0.20}_{-0.14}$ & $0.49^{+0.05}_{-0.04}$ & $\cdots$ & $23.37^{+6.45}_{-5.80}$ & $78.76^{+4.76}_{-4.92}$ & 1.22\\ 
ESO116-G012 & $7.23^{+0.35}_{-0.30}$ & $0.96^{+0.17}_{-0.19}$ & $1.12^{+0.60}_{-0.63}$ & $2.18^{+0.54}_{-0.64}$ & $6.62^{+2.15}_{-2.19}$ & $0.52^{+0.18}_{-0.25}$ & $0.50^{+0.05}_{-0.04}$ & $\cdots$ & $13.38^{+3.44}_{-3.32}$ & $74.09^{+2.71}_{-2.71}$ & 1.06\\
\hline
\end{longtable}
%Table of the marginalized physical values (Vmax, Sigma_Vmax, ...)
%\footnotesize
\fontsize{7pt}{10pt}\selectfont
\begin{longtable}{*{5}{c}}
\caption{The estimated dark matter halo properties of 20 ultrafaint dSphs, 2 ultra diffuse galaxies, the 8 classical dSphs in the Milky Way, 115 SPARC galaxies, and 36 Galaxy Groups \& Galaxy Clusters, marginalized over the resulting model parameters.}
%$V_\mathrm{max}$, $r_{V_\mathrm{max}}$, $\Sigma_\mathrm{DM}(< 0.01r_{V_\mathrm{max}})$, and $\log_{10} (M_\mathrm{baryon}/M_\mathrm{DM})$
\label{tab:Results2}
\hline
%\multicolumn{5}{c}{}\\
Galaxy Name & $V_\mathrm{max}$ & $r_{V_\mathrm{max}}$ & $\Sigma_\mathrm{DM}(< 0.01r_{V_\mathrm{max}})$ & $\log_{10} (M_\ast/M_\mathrm{DM})$ \\
& [km~s$^{-1}$] & [kpc]  &  [$M_\odot$~pc$^{-2}$]& \\ 
\hline
\endfirsthead
\hline
Galaxy Name & $V_\mathrm{max}$ & $r_{V_\mathrm{max}}$ & $\Sigma_\mathrm{DM}(< 0.01r_{V_\mathrm{max}})$ & $\log_{10} (M_\mathrm{baryon}/M_\mathrm{DM})$ \\
& [km~s$^{-1}$] & [kpc]  &  [$M_\odot$~pc$^{-2}$]& \\ 
\hline
\endhead
\hline
\endfoot
\hline
\endlastfoot
\textbf{UFDs \& UDGs} & & & & \\
Antlia 2 & $35.90^{+30.66}_{-14.45}$ & $29.27^{+34.57}_{-15.44}$ & $0.86^{+1.24}_{-0.52}$ & $-3.64^{+0.18}_{-0.16}$\\ 
Bo$\ddot{\mathrm{o}}$tes I & $27.79^{+111.48}_{-16.69}$ & $3.55^{+24.59}_{-3.13}$ & $27.06^{+84.62}_{-20.53}$ & $-4.55^{+0.22}_{-0.21}$\\ 
Canes Venatici I & $38.14^{+96.94}_{-21.24}$ & $8.16^{+31.84}_{-6.73}$ & $18.18^{+25.96}_{-12.78}$ & $-3.90^{+0.19}_{-0.19}$\\ 
Canes Venatici II & $31.75^{+119.73}_{-19.43}$ & $3.54^{+22.08}_{-3.05}$ & $47.63^{+111.68}_{-31.75}$ & $-4.74^{+0.23}_{-0.22}$\\ 
Coma Berenices & $46.51^{+268.00}_{-34.08}$ & $3.30^{+25.29}_{-2.94}$ & $58.53^{+126.45}_{-42.02}$ & $-4.93^{+0.23}_{-0.23}$\\ 
Crater 2 & $7.90^{+16.24}_{-3.34}$ & $7.59^{+28.53}_{-5.90}$ & $0.54^{+2.77}_{-0.40}$ & $-4.05^{+0.20}_{-0.20}$\\ 
Eridanus II & $24.78^{+46.91}_{-9.42}$ & $3.33^{+17.43}_{-2.80}$ & $83.71^{+407.07}_{-52.90}$ & $-4.28^{+0.21}_{-0.20}$\\
\hline
\end{longtable}

\end{document}